\documentclass[letterpaper,english,aps,prb,twocolumn,floatfix,
  showpacs, amsfonts, amssymb]{revtex4} 
\usepackage[T1]{fontenc}
\usepackage[latin1]{inputenc}
\usepackage{amsmath}
\usepackage{graphicx}
\usepackage{amssymb}

\makeatletter

\providecommand{\LyX}{L\kern-.1667em\lower.25em\hbox{Y}\kern-.125emX\@}

\usepackage{amscd}
\usepackage{bm}

\usepackage{babel}
\makeatother
\begin{document}

\title{Domain wall scattering in an interacting one-dimensional electron gas}

\author{R.~G.~Pereira}

\affiliation{Instituto de Física Gleb Wataghin, Unicamp, Caixa Postal 6165, 13083-970 Campinas,
SP, Brazil}

\author{E.~Miranda}

\affiliation{Instituto de Física Gleb Wataghin, Unicamp, Caixa Postal 6165, 13083-970 Campinas,
SP, Brazil}

\date{\today}

\begin{abstract}
We study the transport in a Luttinger liquid coupled to a magnetic chain containing
a Bloch domain wall. We compute the leading correction to the adiabatic limit of
a long domain wall, which causes no scattering. We show that the problem is reminiscent
of an impurity in a Luttinger liquid, but with a different dependence on the interaction
parameters due to spin-flip scattering. For repulsive interactions, we find that
the domain wall resistance diverges with decreasing temperature. This may be relevant
for the design of one-dimensional systems with large magnetoresistance at low temperatures.
\end{abstract}

\pacs{75.47.Jn, 85.75.-d, 73.63.Nm}

\maketitle

The large magnetoresistance associated with the nucleation of domain
walls in magnetic wires and
nanocontacts\cite{garcia,tatara,ebels,imamura,chung,dumpich,dugaev}
has potential applications in the design of high-density magnetic
memories and sensors.  The negative magnetoresistance observed in
these systems was originally explained by the mistracking of carrier
spins when the local magnetization rotates in a distance comparable to
the Fermi wavelength.\cite{cabrera} Strictly speaking, none of the
available experiments has reached the extreme one-dimensional (1D)
limit. It would be interesting to look for effects specific to 1D
systems, whose transport properties are unique. These systems fall
into the universality class of Luttinger Liquids (LL), which are
distinguished by the absence of stable quasi-particle
excitations.\cite{voit} A clear signature of a LL is a power-law
dependence of the conductance through a non-magnetic
impurity.\cite{kanefisher,furusaki,yue} At $T=0$, a vanishingly small
barrier is able to produce perfect reflection if the carriers interact
repulsively.

The effect of non-magnetic impurities suggests a similar phenomenon in
the case of a magnetic inhomogeneity. In this article we show that a
magnetic domain wall behaves as a spin-flip impurity in a LL. We
analyze the backscattering term of the domain wall in the limit of
weak scattering. It is governed by an anomalous dimension given
primarily by $\left(K_{c}+K_{s}^{-1}\right)/2$, where $K_{c}$ and
$K_{s}$ are the LL interaction parameters. There is also a correction
due to the asymmetry between up and down spin electrons introduced by
the exchange field. In the case of local repulsive interactions, this
should lead to an anomalously large and temperature-dependent
magnetoresistance in one-dimensional systems.

We consider interacting electrons coupled to a magnetic domain wall as described
by the Hamiltonian\begin{equation}
H=\sum _{k\sigma }\epsilon _{k}c_{k\sigma }^{\dagger }c_{k\sigma }^{\phantom {\dagger }}-J_{K}\sum _{j}\mathbf{S}_{j}\cdot \mathbf{s}_{j}+H_{int},\label{eq:hamiltonian}\end{equation}
where $c_{k\sigma }$ destroys a conduction electron with momentum $k$ and spin
projection $\sigma $, $\epsilon _{k}=k^{2}/2m$ for quadratic dispersion, $J_{K}$
is the Kondo coupling constant between conduction electrons and localized spins $\mathbf{S}_{j}$,
and $\mathbf{s}_{j}=\frac{1}{2}\sum _{\alpha \beta }c_{j\alpha }^{\dagger }\bm \sigma _{\alpha \beta }c_{j\beta }^{\phantom {\dagger }}$
is the conduction electron spin density at site $j$. We assume a
\emph{static, pinned} magnetic domain wall
described in the continuum limit by setting $\mathbf{S}\left(x\right)=S\cos \theta \left(x\right)\hat{\mathbf{z}}+S\sin \theta \left(x\right)\hat{\mathbf{y}}$.
For a Bloch domain wall, we take $\cos \theta \left(x\right)=-\tanh \left(x/\lambda \right)$,
with $\lambda $ being the wall width. The term $H_{int}$ accounts for electron-electron
interactions. In a uniformly magnetized system, the spin polarization gives rise
to different interaction constants $g_{\uparrow }$ and $g_{\downarrow }$ between
electrons with the same spin  and
$g_{\perp }$ between electrons with opposite spins, due to the absence of
SU$\left(2\right)$ symmetry.\cite{penc-solyom} In a system 
containing a domain wall, the local magnetization acts as an effective magnetic field
on the conduction electrons. The first order approach is to assume that the interaction
constants are different for the spin densities in the direction fixed by the spin
background\[
H_{int}=\int dx\, \left\{ \frac{g_{\uparrow }}{2}\rho _{\wedge }^{2}+\frac{g_{\downarrow }}{2}\rho _{\vee }^{2}+g_{\perp }\rho _{\wedge }\rho _{\vee }\right\} ,\]
where \[
\rho _{\wedge ,\vee }\left(x\right)=\psi ^{\dagger }\left(x\right)\frac{1\pm \bm \sigma \cdot \mathbf{e}\left(x\right)}{2}\psi \left(x\right),\]
and $\mathbf{e}\left(x\right)=\cos \theta \left(x\right)\hat{\mathbf{z}}+\sin \theta \left(x\right)\hat{\mathbf{y}}$.
This expression should be exact in the limit of long domain walls. For low polarizations
($J_{K}\rightarrow 0$), we recover spin degeneracy and all the interaction constants
must be equal ($g_{\uparrow }=g_{\downarrow }=g_{\perp }$).

It is now convenient to perform a spin gauge transformation that aligns the spin
of the conduction electrons with the local magnetization.\cite{tatara} This amounts
to rotating the spin density operator $\mathbf{s}\left(x\right)$ by the angle $\theta \left(x\right)$
around the $x$ axis, which is accomplished by the operator\[
U=\exp \left\{ \frac{i}{2}\int dx\, \theta \left(x\right)\left(\psi _{\downarrow }^{\dagger }\psi _{\uparrow }+\psi _{\uparrow }^{\dagger }\psi _{\downarrow }\right)\right\} ,\]
 where $\psi _{\sigma }\left(x\right)$ is the field operator for conduction electrons.
The rotation of $H$ through $U$ yields\begin{equation}
\tilde{H}=U^{\dagger }HU=\sum _{k\sigma }\epsilon _{k\sigma }c_{k\sigma }^{\dagger }c_{k\sigma }^{\phantom {\dagger }}+\tilde{H}_{int}+H_{w}.\label{eq:transformedhamilt}\end{equation}
Here, $\epsilon _{k\sigma }=\epsilon _{k}-\sigma J_{K}S/2$ expresses the fact that
the effective magnetic field of the local moments breaks the spin degeneracy of the
electron gas. $\tilde{H}_{int}$ is obtained from $H_{int}$ by changing $\rho _{\wedge ,\vee }\rightarrow \rho _{\uparrow ,\downarrow }$.
The transformation also makes explicit the scattering term due to the presence of
the domain wall \begin{equation}
H_{w}=-\frac{i}{4m}\int dx\, \partial _{x}\theta \, \psi ^{\dagger }\sigma _{x}\partial _{x}\psi +H.c.+O\left(\lambda ^{-2}\right),\label{eq:hw}\end{equation}
where we intend to carry out the calculations to leading order in $1/\lambda $.
This corresponds to the first correction to the adiabatic limit of a very long domain
wall, which produces no scattering.

We now focus on the long-wavelength limit of the conduction electrons. In this limit,
we can linearize the dispersion around the Fermi points. Since each of the two spin
branches has a different Fermi wave vector $k_{F\sigma }$, we must have two Fermi
velocities $v_{F\sigma }=v_{F}\left(1+\sigma \zeta \right)$, with $v_{F}$ the mean
Fermi velocity and $\zeta $ the velocity mismatch\[
\zeta =\frac{v_{F\uparrow }-v_{F\downarrow }}{v_{F\uparrow }+v_{F\downarrow }}.\]
The linearized dispersion for spin $\sigma $ reads $\epsilon _{k\sigma }=v_{F\sigma }\left(k\mp k_{F\sigma }\right)$,
where the minus (plus) sign applies to right (left) moving electrons. The field operator
$\psi _{\sigma }$ then naturally separates into right and left parts \[
\psi _{\sigma }\left(x\right)=e^{ik_{F\sigma }x}\psi _{+,\sigma }\left(x\right)+e^{-ik_{F\sigma }x}\psi _{-,\sigma }\left(x\right).\]

Bosonization enables one to build an effective theory by mapping the fermionic operators
into associated bosonic fields.\cite{voit} In terms of these fields, the field operators
are given by\begin{equation}
\psi _{r,\sigma }\left(x\right)=\frac{1}{\sqrt{2\pi \alpha }}\, \exp \left\{ -i\sqrt{\pi }\left[\theta _{\sigma }\left(x\right)-r\phi _{\sigma }\left(x\right)\right]\right\} ,\label{eq:fieldoperator}\end{equation}
where $\alpha ^{-1}$ is a momentum cutoff and $\phi _{\sigma }$ and $\theta _{\sigma }$
are dual fields satisfying $\left[\phi _{\sigma }\left(x\right),\partial _{x}\theta _{\sigma }\left(x^{\prime }\right)\right]=i\delta \left(x-x^{\prime }\right)$.
We further define the charge and spin bosons $\phi _{c,s}=\left(\phi _{\uparrow }\pm \phi _{\downarrow }\right)/\sqrt{2}$.
Upon bosonizing the free part of the Hamiltonian (\ref{eq:transformedhamilt}), we
get the LL Hamiltonian\cite{voit}\begin{eqnarray}
H_{LL} & = & \sum _{\nu =c,s}\frac{v_{\nu }}{2}\int dx\, \left\{ K_{\nu }\left(\partial _{x}\theta _{\nu }\right)^{2}+\frac{1}{K_{\nu }}\left(\partial _{x}\phi _{\nu }\right)^{2}\right\} \nonumber \\
 & + & \int dx\, \left\{ \zeta v_{1}\partial _{x}\theta _{c}\partial _{x}\theta _{s}+\zeta v_{2}\partial _{x}\phi _{c}\partial _{x}\phi _{s}\right\} ,\label{eq:luttinger}
\end{eqnarray}
where \begin{eqnarray*}
v_{1} & = & v_{F}+\frac{g_{4\uparrow }-g_{4\downarrow }+g_{2\uparrow }-g_{2\downarrow }}{2\pi \zeta },\\
v_{2} & = & v_{F}+\frac{g_{4\uparrow }-g_{4\downarrow }-g_{2\uparrow }+g_{2\downarrow }}{2\pi \zeta },
\end{eqnarray*}
where $g_{2\sigma }$ and $g_{4\sigma }$ are the interaction constants between electrons
in different branches and in the same branch, respectively. For not very large $\zeta $,
we will take $g_{i\uparrow }-g_{i\downarrow }\propto \zeta $ ($i=2,4$)\cite{penc-solyom}
so that $v_{1,2}$ are approximately independent of $\zeta $. It is clear from Eq.~(\ref{eq:luttinger})
that the spin background introduces scattering between charge and spin excitations,
which are no longer the normal modes of system.

The bosonized form of the scattering terms $H_{w}$ can be obtained easily by using
the relation (\ref{eq:fieldoperator}). We retain only the backscattering term, which
scatters electrons from right to left moving states (and vice-versa) and is important
for the departure from perfect conductance.\cite{kanefisher,furusaki,yue} It can
be written in terms of charge and spin fields as\begin{equation}
H_{w}^{\left(b\right)}=\frac{\zeta k_{F}A_{2k_{F}}}{m\pi \alpha }\sin \left[\sqrt{2\pi }\theta _{s}\left(0\right)\right]\sin \left[\sqrt{2\pi }\phi _{c}\left(0\right)\right],\label{eq:backscattering}\end{equation}
where $k_{F}=\left(k_{F\uparrow }+k_{F\downarrow }\right)/2$ and $A_{q}=\int dx\, e^{-iqx}\partial _{x}\theta \left(x\right)$
is real for symmetric walls. We note that the $2k_{F}$-mode of the domain wall cancels
the oscillation of the backscattering term. Moreover, the scattering amplitude increases
with growing $\zeta $ and thinner walls. 

The free Hamiltonian $H_{LL}$ as given by (\ref{eq:luttinger}) is not in diagonal
form. However, it is still quadratic in the bosonic fields and can be diagonalized
by means of a canonical transformation to new fields $\theta _{c,s}^{\prime }$ and
$\phi _{c,s}^{\prime }$. We define the bosonic field vectors\[
\theta =\left(\begin{array}{c}
 \theta _{c}\\
 \theta _{s}\end{array}\right)\qquad ,\qquad \phi =\left(\begin{array}{c}
 \phi _{c}\\
 \phi _{s}\end{array}\right),\]
so that $H_{LL}$ can be rewritten as\[
H_{LL}=\frac{1}{2}\int dx\, \left\{ \partial _{x}\theta A\partial _{x}\theta +\partial _{x}\phi B\partial _{x}\phi \right\} ,\]
where we have introduced the matrices\[
A=\left(\begin{array}{cc}
 v_{c}K_{c} & \zeta v_{1}\\
 \zeta v_{1} & v_{s}K_{s}\end{array}\right)\quad ,\quad B=\left(\begin{array}{cc}
 v_{c}/K_{c} & \zeta v_{2}\\
 \zeta v_{2} & v_{s}/K_{s}\end{array}\right).\]

Our aim is to diagonalize $A$ and $B$ simultaneously. In order for the LL to be
stable, the corresponding eigenvalues (the velocities of the natural excitations)
must be positive; this limits the validity of our solution to the interval\begin{equation}
\frac{\zeta ^{2}v_{1}^{2}}{v_{c}v_{s}}<K_{c}K_{s}<\frac{v_{c}v_{s}}{\zeta ^{2}v_{2}^{2}}.\label{eq:restriction}\end{equation}
Outside this interval, the polarization is large enough to make one of the velocities
vanish and the spinon-like excitation becomes gapped. We start the diagonalization
by rotating $A$ and $B$ through an angle $\varphi $, as expressed by the matrix\[
R=\left(\begin{array}{cc}
 \cos \varphi  & \sin \varphi \\
 -\sin \varphi  & \cos \varphi \end{array}\right).\]
We choose the angle $\varphi $ in such a way that, applying next the rescaling \[
\Lambda =\left(\begin{array}{cc}
 \sqrt{\kappa } & 0\\
 0 & \sqrt{\mu }\end{array}\right),\]
we shall have $\Lambda ^{-1}R^{t}AR\Lambda ^{-1}=\Lambda R^{t}BR\Lambda $. This
condition requires\begin{subequations}\label{kapamu}

\begin{eqnarray}
\kappa  & = & \sqrt{\frac{v_{c}K_{c}\cos ^{2}\varphi -\zeta v_{1}\sin 2\varphi +v_{s}K_{s}\sin ^{2}\varphi }{\frac{v_{c}}{K_{c}}\cos ^{2}\varphi -\zeta v_{2}\sin 2\varphi +\frac{v_{s}}{K_{s}}\sin ^{2}\varphi }},\label{eq:kapa}\\
\mu  & = & \sqrt{\frac{v_{s}K_{s}\cos ^{2}\varphi +\zeta v_{1}\sin 2\varphi +v_{c}K_{c}\sin ^{2}\varphi }{\frac{v_{s}}{K_{s}}\cos ^{2}\varphi +\zeta v_{2}\sin 2\varphi +\frac{v_{c}}{K_{c}}\sin ^{2}\varphi }},\label{eq:mu}\\
\kappa \mu  & = & \frac{2\zeta v_{1}\cos 2\varphi +\left(v_{c}K_{c}-v_{s}K_{s}\right)\sin 2\varphi }{2\zeta v_{2}\cos 2\varphi +\left(\frac{v_{c}}{K_{c}}-\frac{v_{s}}{K_{s}}\right)\sin 2\varphi }.\label{eq:kapaXmu}
\end{eqnarray}
\end{subequations}The restriction (\ref{eq:restriction}) assures that $\kappa $
and $\mu $ are both real. We then determine $\varphi $ in the interval $\left[-\pi /4,\pi /4\right]$
for arbitrary $\zeta $ by imposing that the three expressions (\ref{kapamu}) are
solved simultaneously. Being equal, the two transformed matrices can be made diagonal
by performing a second rotation $S$. As a result, the Hamiltonian (\ref{eq:luttinger})
assumes the form \begin{equation}
H_{LL}=\sum _{\nu =c,s}\frac{v_{\nu }^{\prime }}{2}\int dx\left\{ \left(\partial _{x}\theta _{\nu }^{\prime }\right)^{2}+\left(\partial _{x}\phi _{\nu }^{\prime }\right)^{2}\right\} ,\label{eq:hamiltdiagonal}\end{equation}
where $v_{c,s}^{\prime }$ are the eigenvalues of the final matrix. The original
bosonic field vectors are written in terms of the new ones as\[
\theta =T^{\theta }\theta ^{\prime }\qquad \qquad \phi =T^{\phi }\phi ^{\prime },\]
where $T^{\theta }=R\Lambda ^{-1}S$ and $T^{\phi }=R\Lambda S=\left[\left(T^{\theta }\right)^{-1}\right]^{t}$.

In order to analyze the effect of the backscattering term (\ref{eq:backscattering}),
we work out an effective action for the free Hamiltonian that depends only on the
fields at the origin.\cite{kanefisher} In terms of the new bosonic fields, we have\begin{eqnarray*}
H_{w}^{(b)} & = & \gamma \sin \left[\sqrt{2\pi }\left(T_{21}^{\theta }\theta _{c}^{\prime }\left(0\right)+T_{22}^{\theta }\theta _{s}^{\prime }\left(0\right)\right)\right]\times \\
 & \times  & \sin \left[\sqrt{2\pi }\left(T_{11}^{\phi }\phi _{c}^{\prime }\left(0\right)+T_{12}^{\phi }\phi _{s}^{\prime }\left(0\right)\right)\right],
\end{eqnarray*}
where $\gamma =\zeta k_{F}A_{2k_{F}}/m\pi \alpha $. Thus, the effective action must
depend on both conjugate fields. We start with the free partition function in imaginary
time\begin{eqnarray*}
Z_{0} & = & \int \prod _{\nu }D\phi _{\nu }^{\prime }D\theta _{\nu }^{\prime }\times \\
 & \times  & \exp \left\{ \int dxd\tau \, \left[i\partial _{\tau }\phi _{\nu }^{\prime }\, \partial _{x}\theta _{\nu }^{\prime }-\mathcal{H}\left(\phi _{\nu }^{\prime },\theta _{\nu }^{\prime }\right)\right]\right\} ,
\end{eqnarray*}
where $\mathcal{H}\left(\phi _{\nu }^{\prime },\theta _{\nu }^{\prime }\right)$
is the Hamiltonian density in (\ref{eq:hamiltdiagonal}). Then we integrate out the
degrees of freedom for $x\neq 0$ and find the effective action\begin{eqnarray*}
S_{0}^{eff}\left[\phi _{0},\theta _{0}\right] & = & \frac{1}{\beta }\sum _{\nu ,n}\left|\omega _{n}\right|\phi _{0\nu }^{\prime }\left(\omega _{n}\right)\phi _{0\nu }^{\prime }\left(-\omega _{n}\right)+\\
 & + & \frac{1}{\beta }\sum _{\nu ,n}\left|\omega _{n}\right|\theta _{0\nu }^{\prime }\left(\omega _{n}\right)\theta _{0\nu }^{\prime }\left(-\omega _{n}\right),
\end{eqnarray*}
where $\omega _{n}$ are bosonic Matsubara frequencies. A renormalization group analysis
gives the flow of the coupling constant $\gamma $ at low energies ($\ell \to \infty $)\cite{kanefisher,furusaki,yue}\[
\frac{d\gamma }{d\ell }=\left(1-D\right)\gamma \]
where $D$ is the dimension of the backscattering operator, given by\begin{equation}
D=\frac{1}{2}\left[\left(T_{21}^{\theta }\right)^{2}+\left(T_{22}^{\theta }\right)^{2}+\left(T_{11}^{\phi }\right)^{2}+\left(T_{12}^{\phi }\right)^{2}\right].\label{eq:dimension}\end{equation}

We would like to express $D$ in terms of the LL parameters. Remarkably, it does
not depend on the matrix $S$ and reduces to\[
D=\frac{1}{2}\left[\kappa \cos ²\varphi +\frac{1}{\kappa }\sin ²\varphi +\frac{1}{\mu }\cos ²\varphi +\mu \sin ²\varphi \right].\]
For small $\zeta $, we get \begin{widetext}\begin{equation}
D=\frac{1}{2}\left(K_{c}+\frac{1}{K_{s}}\right)+\frac{\left(K_{c}K_{s}v_{2}-v_{1}\right)\left[K_{c}K_{s}\left(K_{c}K_{s}v_{2}+v_{1}\right)v_{s}^{2}+2\left(K_{c}^{2}K_{s}^{2}v_{2}-v_{1}\right)v_{c}v_{s}-(K_{c}K_{s}v_{2}+v_{1})v_{c}^{2}\right]\zeta ²}{4K_{c}K_{s}^{2}v_{c}v_{s}\left(v_{c}+v_{s}\right)^{2}}.\label{eq:dexpansion}\end{equation}
\end{widetext} For non-magnetic impurities, $D_{imp}=\left(K_{c}+K_{s}\right)/2$,
which is different from the $\zeta \rightarrow 0$ limit of our result. This should
be attributed to the spin-flip scattering explicit in the form (\ref{eq:hw}), in
contrast with the charge-only scattering by a non-magnetic impurity. 

\begin{figure}
\begin{center}\includegraphics[  scale=0.3]{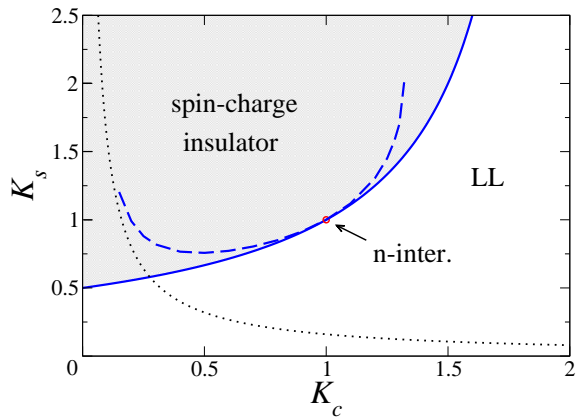}\end{center}

\caption{\label{phase-diagram}Phase diagram for a Luttinger Liquid coupled to a magnetic
domain wall. The backscattering term $H_{w}^{\left(b\right)}$ is marginal on the
straight line in the limit $\zeta \rightarrow 0$, and on the dashed one for $\zeta =0.4$
(with all velocities equal). The dotted line corresponds to the lower bound of stability
of the Luttinger Liquid according to equation (\ref{eq:restriction}).}
\end{figure}

The possible phases can be obtained similarly to Refs.~\onlinecite{kanefisher,furusaki,yue}.
We first focus on the $\zeta \rightarrow 0$ case. For $D>1$ or $K_{c}+K_{s}^{-1}>2$,
the scattering is irrelevant and the fixed point is a LL with perfect transmission
of charge and spin. For $D<1$ or $K_{c}+K_{s}^{-1}<2$, which is favored for increasingly
repulsive interactions (decreasing $K_{c}$), the scattering is relevant and the
system flows to the strong coupling limit. This limit corresponds to two semi-infinite
LLs with spins polarized in opposite directions and coupled through a small hopping
term that flips the electron spin in the tunneling process. This term has been analyzed
in the context of a magnetic impurity in a LL \cite{fabrizio}, where the hopping
is found to be irrelevant for repulsive interactions. As a result, the fixed point
is a spin-charge insulator at $T=0$. The straight line in Fig.~\ref{phase-diagram}
represents the marginal line $D=1$ in the limit $\zeta \rightarrow 0$. %

The correction for finite $\zeta $ vanishes when $K_{c}K_{s}=v_{1}/v_{2}$. Actually,
this cancellation happens to all orders in $\zeta $ because the equations (\ref{kapamu})
are always satisfied for $\varphi =0$, $\kappa =K_{c}$ and $\mu =K_{s}=v_{1}/v_{2}K_{c}$.
Consequently, the condition $K_{c}K_{s}=v_{1}/v_{2}$ defines a line in parameter
space where the dimension of the scattering term is $\zeta $-invariant. In particular,
the non-interacting point $K_{c}=K_{s}=1$ (and $v_{1}=v_{2}=v_{F}$) is always marginal.
For $K_{c}K_{s}\neq v_{1}/v_{2}$, the dimension varies with $\zeta $. The dashed
line in Fig.~\ref{phase-diagram} shows how the marginal line is modified for $\zeta =0.4$
and $v_{c}=v_{s}=v_{1}=v_{2}=v_{F}$.

The dimension $D$ manifests itself in the exponent of the frequency-dependent domain
wall resistance. The resistivity associated with the backscattering off the wall
at low frequencies is $\rho \left(\omega \right)\propto \omega ^{2\left(D-1\right)}$.
Likewise, the finite-temperature resistance turns out to be $\rho \left(T\right)\propto T^{2\left(D-1\right)}$.
Therefore, the domain wall scattering in a LL gives rise to a temperature-dependent
resistance. For $D>1$, the resistance vanishes as a power law when $T\rightarrow 0$;
for $D<1$, it diverges in the limit $T\rightarrow 0$. The LL behavior is cut off
at a temperature $T^{*}\sim v_{F}/L$, below which the transport is dominated by
the Fermi Liquid leads.\cite{kanefisher} This can be understood as follows. The
domain wall is known to induce long-ranged spin density oscillations in the electron
gas.\cite{dugaev} Similarly to what happens with charge density oscillations created
by non-magnetic impurities,\cite{yue} the scattering by these spin density oscillations
diverges at low $\omega $ in one dimension. As a result, the electrons are totally
reflected by the wall.

Finally, let us estimate the exponent in the particular case of the Hubbard model.\cite{penc-solyom}
Due to the absence of SU(2) symmetry, we cannot take $K_{s}=1$ as usual. Instead,
the parameters $K_{c}$ and $K_{s}$ depend implicitly on $\zeta $. To lowest order
in $\zeta $, $K_{s}\approx 1+\left[2\ln \left(\zeta ^{-1}\right)\right]^{-1}$.
Note that this correction has a lower order dependence on $\zeta $ than the explicit
one (order $\zeta ^{2}$) in Eq.~(\ref{eq:dexpansion}). Furthermore, a finite polarization
makes $K_{s}>1$ and so pushes the model into the insulating region of the phase
diagram. For small $U$, $K_{c}\approx 1-aU/2\pi v_{F}+O\left(\zeta \right)$, where
$U$ is the on-site repulsion and $a$ is the lattice spacing. Then, $D\approx 1-aU/4\pi v_{F}-\left[4\ln \left(\zeta ^{-1}\right)\right]^{-1}$.
As an experimental test of this theory, one should look for the dependence of the
resistance exponent on the polarization $\zeta $ of the underlying system of carriers.

In conclusion, we have shown that the domain wall scattering in a Luttinger Liquid
is the magnetic analogue of the Kane-Fisher problem. Just as a non-magnetic impurity,
a domain wall breaks the translation symmetry of the electron gas. The $2k_{F}$
mode of the wall gives rise to a spin-flip backscattering term which is relevant
for repulsive interactions. In this case, the magnetoresistance diverges as a power
law in the limit of zero temperature. By applying magnetic fields one can insert
or remove a single domain wall and then switch between a spin-charge insulator and
a Luttinger Liquid with perfect conductance. This should be relevant in view of the
quest for systems exhibiting large magnetoresistance.

This work was supported by Fapesp through grants 01/12160-5 (RGP) and 01/00719-8
(EM), and by CNPq through grant 301222/97-5 (EM).

\end{document}